# Astronomical Infrared Spectrum of Planetary Nebula Lin49 and Tc1 Identified by Ionized Polycyclic-Pure-Carbon C23 and C60

NORIO OTA

Graduate School of Pure and Applied Sciences, University of Tsukuba,
1-1-1 Tenoudai Tsukuba-city 305-8571, Japan;   n-otajitaku@nifty.com

Astronomical dust molecule of carbon-rich nebula-Lin49 and nebula-Tc1 could be identified to be polycyclic-pure-carbon C23 by the quantum-chemical calculation. Two driving forces were assumed. One is high speed proton attack on coronene-$C_{24}H_{12}$, which created void-induced $C_{23}H_{12}$. Another is high energy photon irradiation, which brought deep photo-ionization and finally caused dehydrogenation to be C23. Infrared spectrum calculation show that a set of ionized C23 (neutral, mono, and di-cation) could reproduce observed many peaks of 28 bands at wavelength from 6 to 38 micrometer. Previously predicted neutral fullerene-C60 could partially reproduce observed spectrum by 5 bands. Also, we tried calculation on ionized-C60, which show fairly good coincidence with observed 10 bands.

Key words: Astro-chemistry, C60, C23, Infrared spectrum, Lin 49 nebula, Tc 1 nebula

## 1, INTRODUCTION

Interstellar infrared spectrum (IR) due to polycyclic aromatic hydrocarbon [PAH] (Boersma et al. 2013, 2014) and polycyclic pure carbon [PPC] (Cami et al. 2010, Otsuka et al., 2016) were observed in many astronomical dust clouds. Identification of particular molecule is essentially important to search chemical evolution step of organics and to study material building block of creation of life in the universe. In 2014, by the first principles quantum calculation, it was found that void induced hydrocarbon molecule $(C_{23}H_{12})^{2+}$ shows very similar infrared IR with interstellar observed one (Ota 2014, 2015). This molecule contains two hydrocarbon pentagons combined with five hexagons. We assumed two astronomical situations. One is high speed particle (mainly proton) attack on a molecule to create a carbon void bringing quantum deformation by the Jahn-Teller effect (Jahn and Teller 1937). Another is high energy photon irradiation on a molecule bringing deep cationic state on hydrocarbon, and finally resulting dehydrogenation, that is, creation of PPC molecule. Through such theoretical approach, we could identify several molecules as like PAHs of $(C_{12}H_8)^{n+}$, $(C_{23}H_{12})^{n+}$, $(C_{53}H_{18})^{n+}$, $(C_3H_2)^{n+}$, and PPCs of $(C_{12})^{n+}$, $(C_{23})^{n+}$, $(C_{53})^{n+}$, $(C_3)^{n+}$ applying to many observed IRs including Herbig Ae young stars (Acke et al. 2010, Ota 2017e) and several Galaxies (Ota 2017c). Detailed identification and quantum-chemical explanation was studied in these two years (Ota 2017a, 2017b, 2017c, 2017d, 2017e, 2018a, 2018b, 2018c, 2018d). Figure 1 show previously identified molecules under above mentioned two principles, that is, (1) void induced quantum deformation and (2) photo-ionized dehydrogenation.

Here, we would like to try coronene-$(C_{24}H_{12})$ based molecular evolution, finally resulting $(C_{23})^{n+}$. Calculated infrared spectrum of $(C_{23})^{n+}$ will be compared with observed IR of carbon-rich and hydrogen-poor planetary nebula Tc1 in the Milky Way galaxy and Lin49 in the Small Magellanic cloud. In a previous study by Cami et al., IR of Tc1 was indicated to be similar with that of carbon fullerene-C60 and C70 (Cami et al. 2010). Also, Otsuka et al. studied detailed IR feature of Lin 49 (Otsuka et al. 2016).

In this study, it will be suggested that a set of IR of $(C_{23})^{n+}$ (n=0, +1, and +2) can reproduce detailed spectrum of Tc1 and Lin49 very well. Also, it will be announced that ionized fullerene $(C_{60})^{n+}$ (n=0, +1 and +2) can reproduce them fairly well.

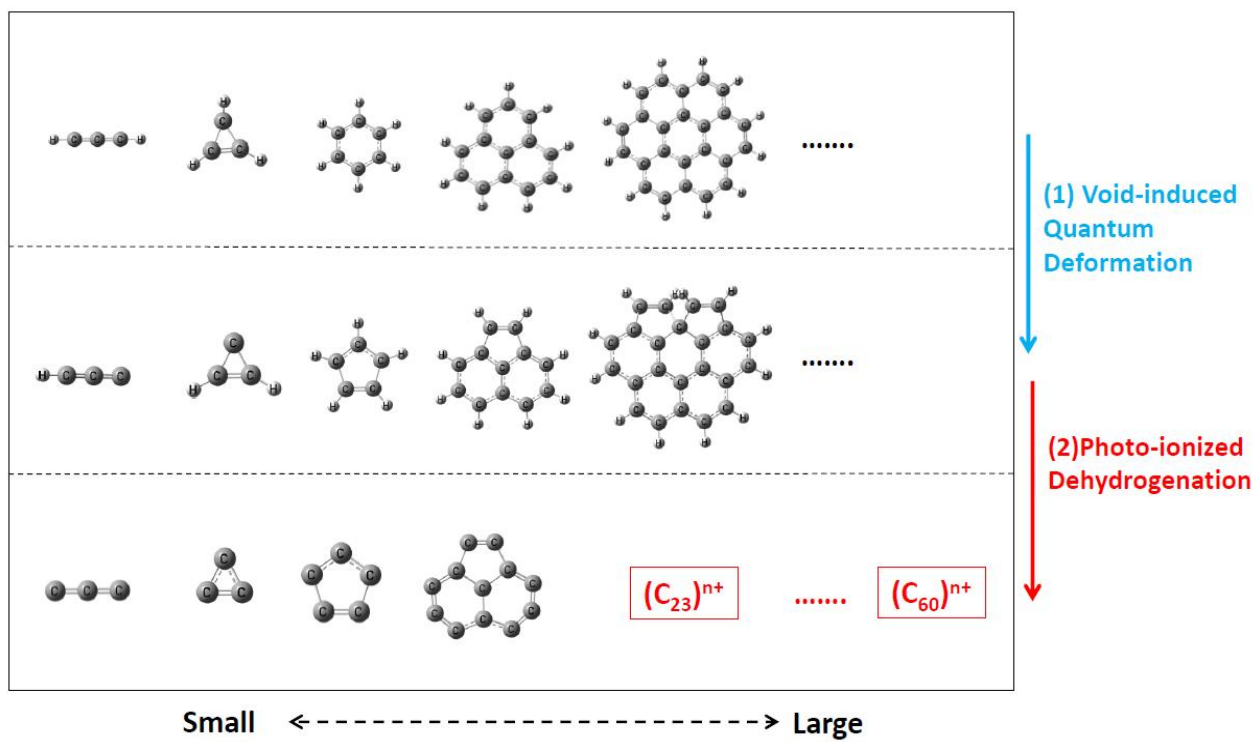

Figure 1, Quantum-chemically identified molecules by two principles, that is, (1) void induced quantum deformation and (2) photo-ionized dehydrogenation. In this study, starting molecule was coronene-($C_{24}H_{12}$) as shown on top right. Void induced-($C_{23}H_{12}$) will be ionized very deeply and dehydrogenated to be ($C_{23}$).

2, CALCULATION METHOD

In quantum chemistry calculation, we have to obtain total energy, optimized atom configuration, and infrared vibrational mode frequency and strength depend on a given initial atomic configuration, charge and spin state Sz. Density functional theory (DFT) with unrestricted B3LYP functional was applied utilizing Gaussian09 package (Frisch et al. 1984, 2009) employing an atomic orbital 6-31G basis set. The first step calculation is to obtain the self-consistent energy, optimized atomic configuration and spin density. Required convergence on the root mean square density matrix was less than $10^{-8}$ within 128 cycles. Based on such optimized results, harmonic vibrational frequency and strength was calculated. Vibration strength is obtained as molar absorption coefficient ε (km/mol.). Comparing DFT harmonic wavenumber $N_{DFT}$ (cm$^{-1}$) with experimental data, a single scale factor 0.965 was used (Ota 2015). Concerning a redshift for the anharmonic correction, in this paper we did not apply any correction to avoid over estimation in a wide wavelength representation from 2 to 30 micrometer.
Corrected wave number N is obtained simply by N (cm$^{-1}$) = $N_{DFT}$ (cm$^{-1}$) x 0.965.
Wavelength λ is obtained by λ (micrometer) = 10000/N(cm$^{-1}$).
Reproduced IR spectrum was illustrated in a figure by a decomposed Gaussian profile with full width at half maximum FWHM=4cm$^{-1}$.

3, VOID CREATION AND PHOTO-IONIZATION

In order to find out particular molecule to reproduce observed infrared spectrum, we applied the top down molecular evolution hypothesis, that is, PAHs would be destructed from larger one to smaller one due to two ubiquitously existing energy sources. One is high speed proton, and another high energy photon. Figure 2 show an example of coronene-$(C_{24}H_{12})$. High speed proton will be ejected from the central star and may sputter carbon site to create a void. Immediately, by the quantum Jahn-Teller effect (1937), there occur quantum deformation on void induced molecule-$(C_{23}H_{12})$, which creating a molecule with hydrocarbon two pentagons combined with five hexagons. After that, photon from the central star will illuminate $(C_{23}H_{12})$ to pull out electrons one by one.

(1) Photo-ionization of PAH
Photo-ionized configuration change of PAH was calculated as shown in Figure 3. Starting molecule is neutral-$(C_{23}H_{12})^{0+}$ pictured at top left, where blue arrow shows electric dipole. Molecular structure having two hydrocarbon pentagons combined with five hexagons were kept up to fifth step of $(C_{23}H_{12})^{5+}$. At sixth ionization of $(C_{23}H_{12})^{6+}$, two-pentagon structure disappears by loosening carbon bonding. At eighth, one hydrogen atom was decoupled from the molecule and single pentagon was newly created. At tenth, there occur complete dehydrogenation, which resulting a super molecule of $(C_{23})$ and released 12H.

(2) Photo-ionization of PPC
Focusing on $(C_{23})$, continued photo-ionization step was calculated as illustrated in Figure 4. Molecular configuration of single carbon pentagon was kept from neutral $(C_{23})^{0+}$ until ninth step of $(C_{23})^{9+}$. Decoupling of carbon to carbon atom had started at tenth step. There occurs photo-dissociation of PPC molecule.

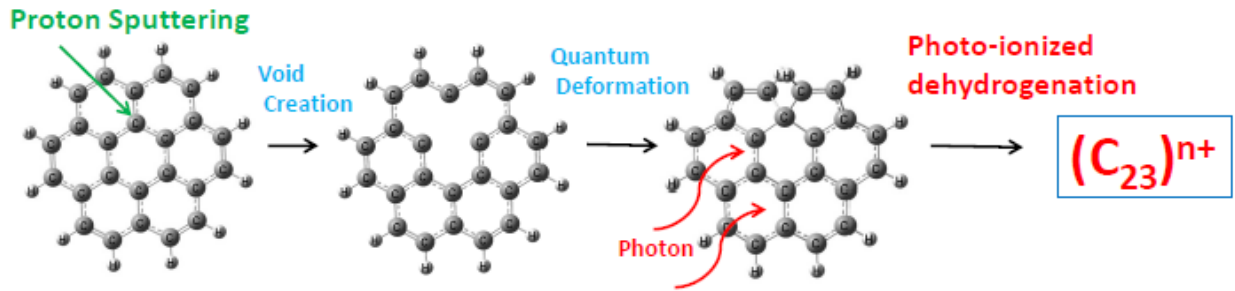

Figure 2, Top down molecular evolution model of coronene-$(C_{24}H_{12})$. High speed proton will sputter carbon site and create a void. Immediately, there occurs quantum deformation as $(C_{23}H_{12})$ resulting a particular molecule having hydrocarbon two pentagons combined with five hexagons. After that, high energy photon from the central star irradiates $(C_{23}H_{12})$ to pull out electrons. Deeply photo-ionized molecule will be dehydrogenated and transformed to pure carbon $(C_{23})$.

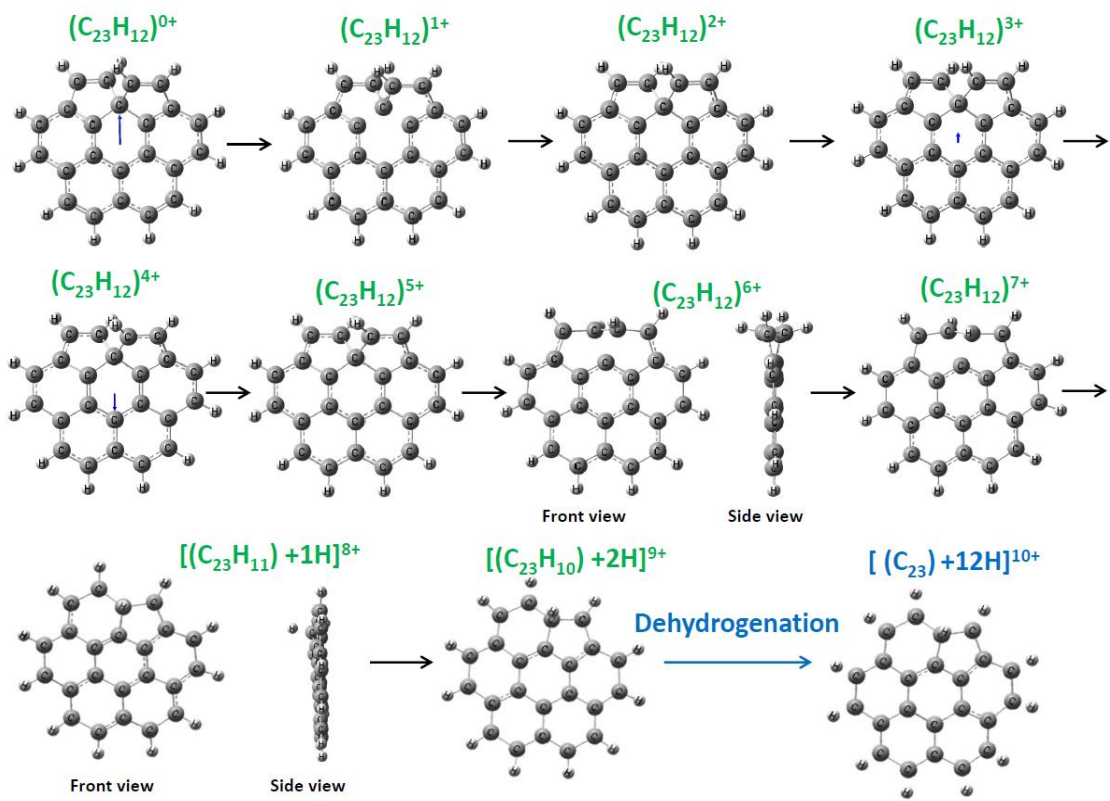

Figure 3, Photo-ionization of polycyclic aromatic hydrocarbon $(C_{23}H_{12})^{n+}$.

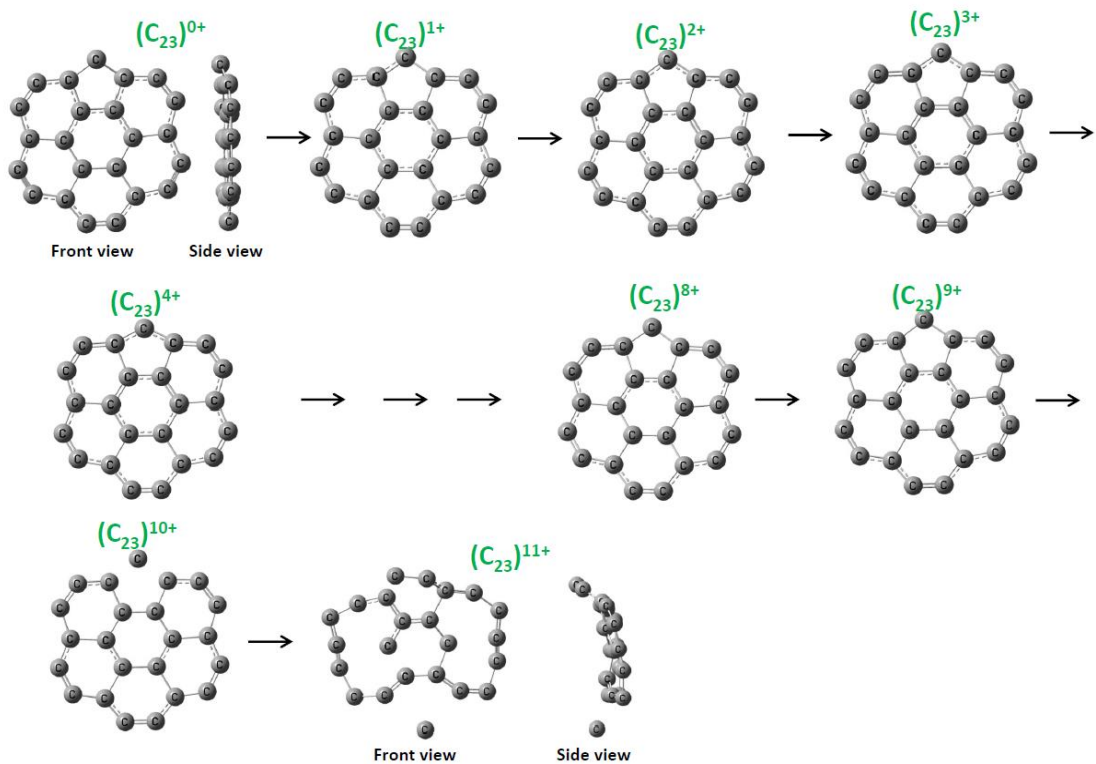

Figure 4, Photo-ionization of polycyclic pure carbon $(C_{23})^{n+}$

4, INFRARED SPECTRUM OF NEBULA-Tc1 AND -Lin49 IDENTIFIED BY $(C_{23})^{n+}$

Infrared spectrum of $(C_{23}H_{12})^{n+}$ and $(C_{23})^{n+}$ were calculated. We already reported that hydrocarbon $(C_{23}H_{12})^{n+}$ could reproduce observed ubiquitous one (Ota 2014, 2017a, 2017b, 2017c). We also expected that carbon molecule $(C_{23})^{n+}$ could reproduce carbon-rich and hydrogen-poor planetary nebula. Famous examples are Tc 1-nebula in the Milky Way galaxy and Lin 49-nebula in the Small Magellanic cloud. In 2010, J. Cami et al. could find out good coincidence of observed IR of Tc 1 with pure carbon fullerene-$(C_{60})$ and $(C_{70})$ at wavelength of 18.9, 17.4, 8.5, and 7.0 micrometer (Cami et al. 2010). Also, M. Otsuka et al. compared observed IR of Lin 49 and Tc 1 as shown on top of Figure 5.

Calculated IR of $(C_{23})^{0+}$, $(C_{23})^{1+}$, and $(C_{23})^{2+}$ were illustrated in Figure 5 comparing with observed one of Lin 49 and Tc 1. It was a surprise that major observed bands could be reproduced by a combination of these three cationic IR at 19.1, 18.8, 17.3, 8.5, and 7.0 micrometer. Large observed peak at 12.8 micrometer may come from atomic emission of Ne II. Also, other calculated bands of 7.5, 10.0, 12.2, 13.2, 17.2, 22.2, 23.2, 29.0, and 36.0 micrometer could be coincide with observed peaks.

Other detailed bands could be identified as shown in Figure 6. Total identified number of bands were 28. It should be noted that zoomed calculated spectrum from 5 to 16 micrometer could reproduce complex observed spectrum very well.

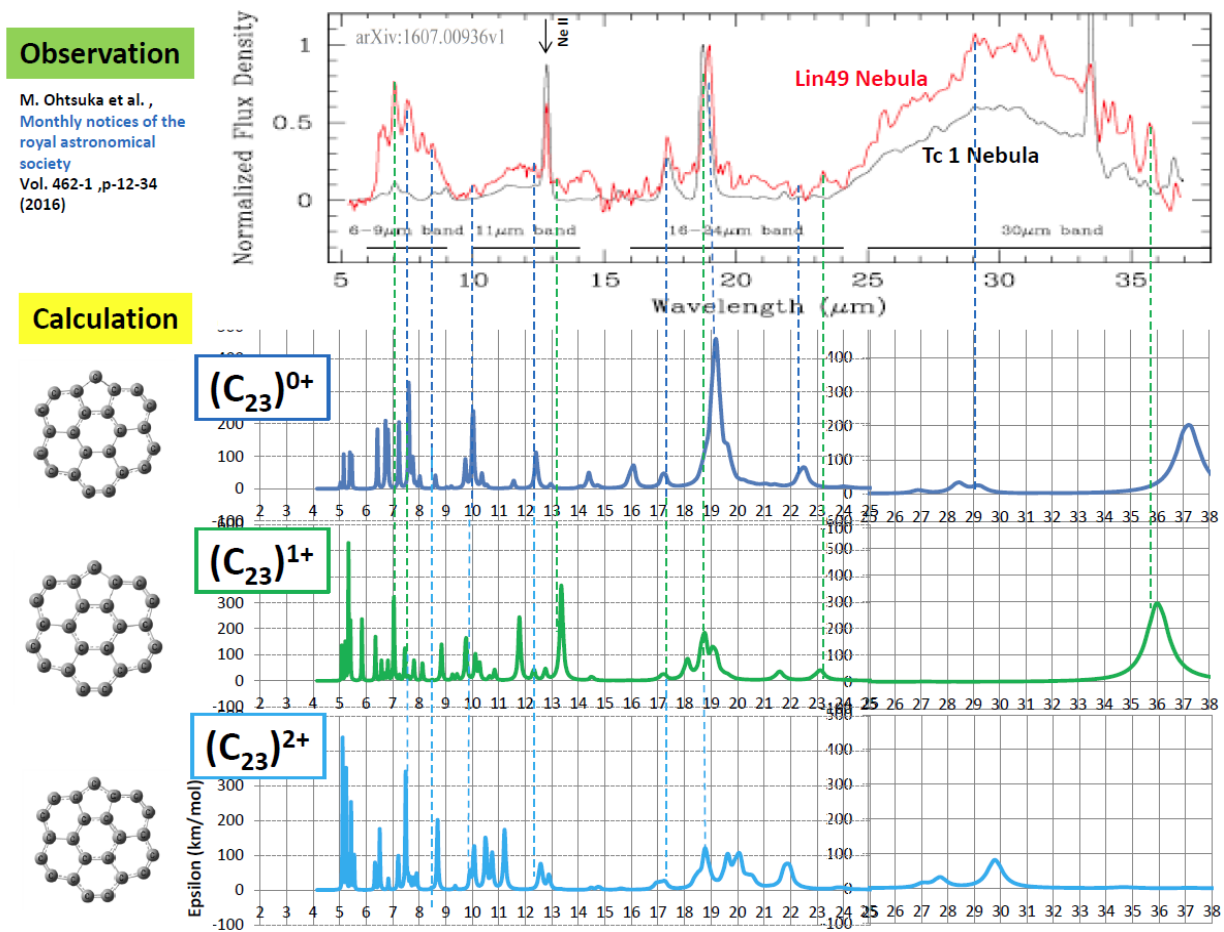

Figure 5, Observed IR of Lin 49 nebula and Tc 1 nebula studied by M. Otsuka et al. (2016) on top panel. A set of calculated IR of $(C_{23})^{0+}$, $(C_{23})^{1+}$, and $(C_{23})^{2+}$ could reproduce observed major bands very well.

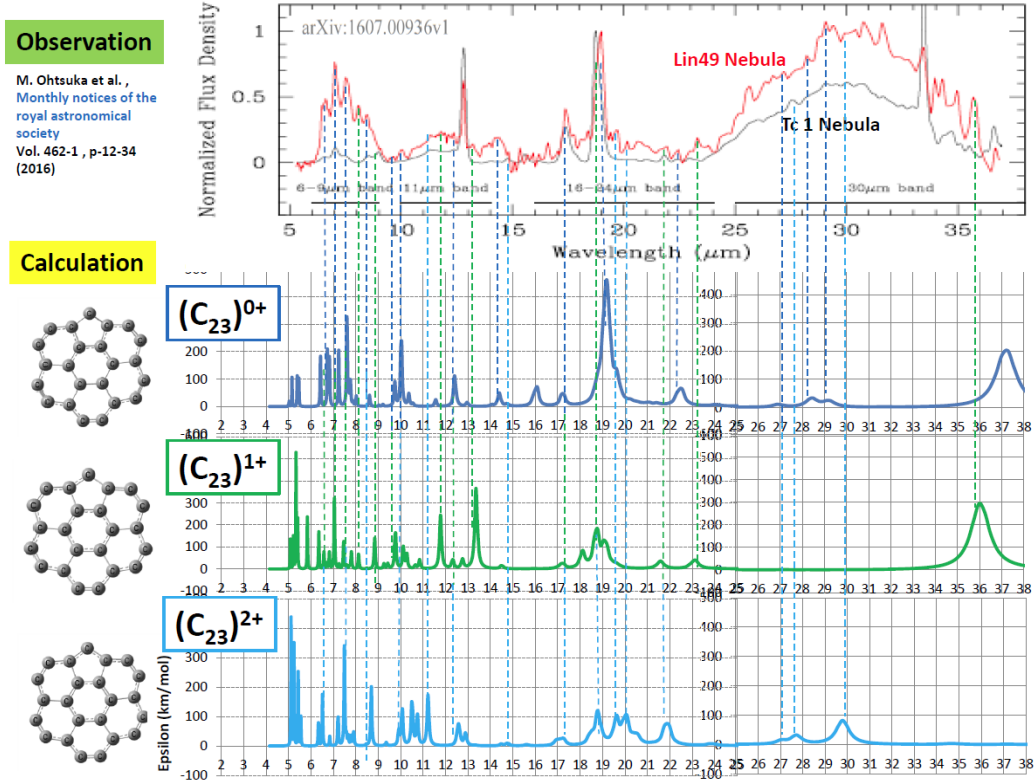

Figure 6, Observed IR of Lin 49 and Tc 1 could be identified by a detailed combination of calculated one of $(C_{23})^{0+}$, $(C_{23})^{1+}$, and $(C_{23})^{2+}$. There are identified 28 bands.

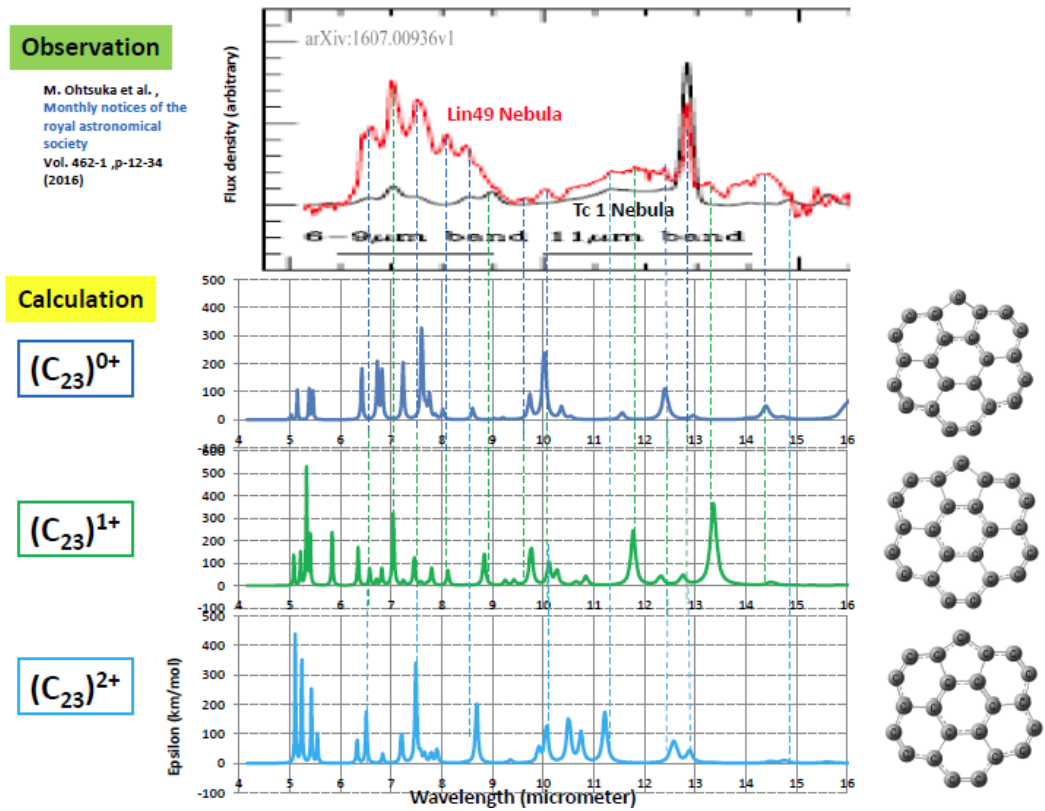

Figure 7, Zoom up of identification from 6 to 16 micrometer.

## 5, CALCULATED IR OF $(C_{60})^{0+}$, $(C_{60})^{1+}$, and $(C_{60})^{2+}$

We should study the capability of previously predicted molecule fullerene-$C_{60}$ (Cami et al. 2010). As shown in Figure 8, calculated IR of neutral-$(C_{60})$ were 18.9, 17.6, 8.7, and 6.8 micrometer, which coincide fairly well with observed one of 19.0, 18.8, 17.4, 8.5, and 7.0 micrometer. It should be noted that calculated IR of cation $(C_{60})^{1+}$ and $(C_{60})^{2+}$ can reproduce other observed bands of 6.6, 7.5, 8.1, 13.2, and 29.1 micrometer. Especially, observed band at 25.6 micrometer was reproduced by $(C_{60})^{1+}$ and $(C_{60})^{2+}$, which bands could not be calculated by a series of $(C_{23})^{n+}$.

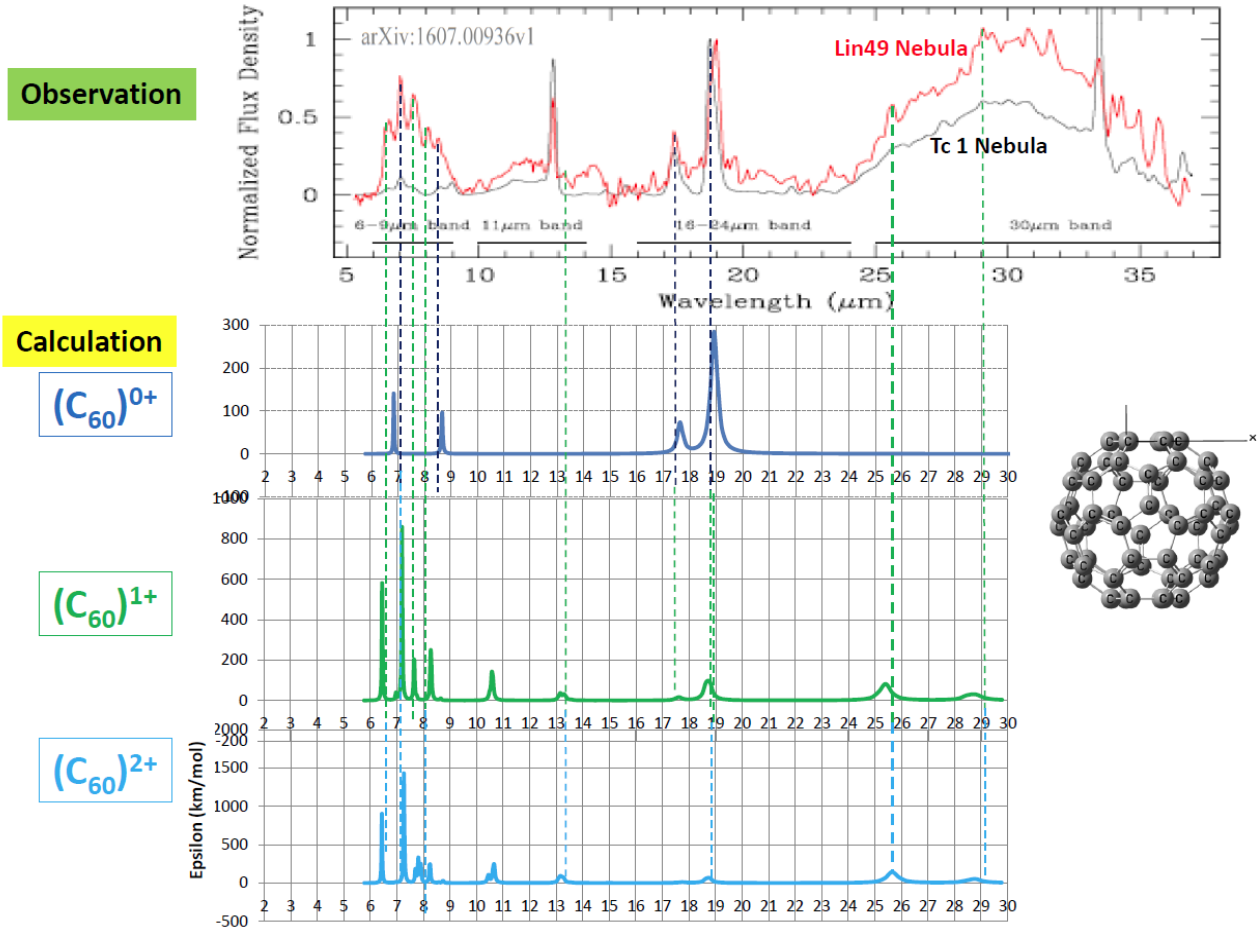

Figure 8, Calculated IR of $(C_{60})^{0+}$, $(C_{60})^{1+}$, and $(C_{60})^{2+}$. Combination of these three spectra can reproduce observed bands by 10 peaks.

## 6, DISCUSSION

We should discuss which molecule is the right carrier of observed IR. As shown in Figure 9 (A), a series of $(C_{23})^{n+}$ (n=0, +1, and +2) could reproduce 28 observed bands very well as marked by black lines. Whereas, as shown in (B), IR of experimentally certified neutral-$(C_{60})^{0+}$ [green line] and $(C_{70})^{0+}$ [blue line] fit at 9 bands (Cami et al. 2010). Comparing with $(C_{23})^{n+}$, number of bands are few. Fullerene-$(C_{60})$ is structured by carbon pentagons combined with hexagons, which is essentially similar with $(C_{23})$. However, the molecular symmetry of $(C_{60})$ is extremely high and many molecular vibrational modes (bands) are canceled each other, which is the reason why larger carbon number of $(C_{60})$ show a fewer active bands. In case of ionized-$(C_{60})$, we can see additional bands as compared in (C). We should pay attention on 25.6 micrometer band reproduced by $(C_{60})^{1+}$ and $(C_{60})^{2+}$, where nothing in case of $(C_{23})^{n+}$. Looking again at observed IR of Lin 49, there are many peaks from 30 to 35 micrometer, but no identified one by both $(C_{23})^{n+}$ and $(C_{60})^{n+}$. Longer wavelength identification should be studied by other candidates.

As illustrated in Figure 9, infrared telescope may observe piled spectrum coming from neutral and ionized $(C_{23})$, some (PPC)…(C60), (C70) and so on.

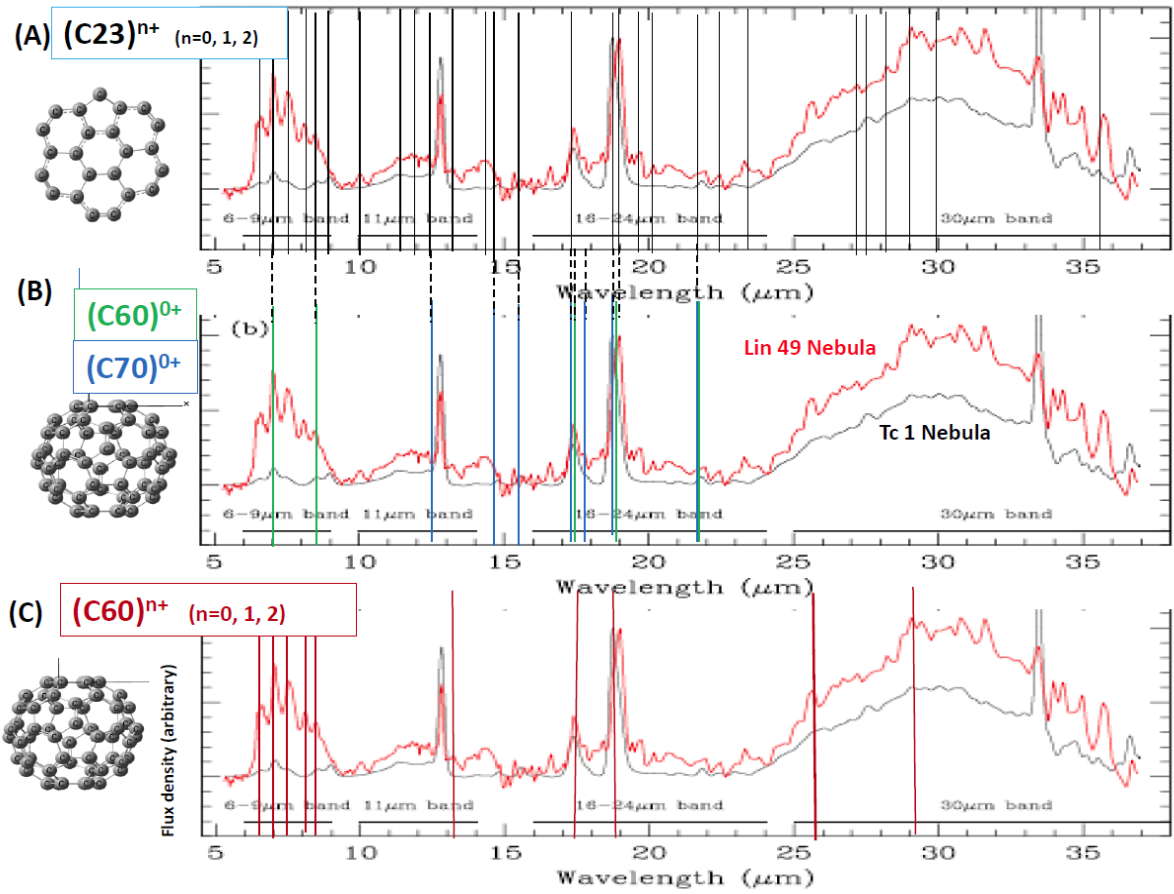

Figure 9, Identified bands in case of $(C_{23})^{n+}$ illustrated in (A), neutral $(C_{60})$ and $(C_{70})$ in (B), and ionized-$(C_{60})^{n+}$ in (C).

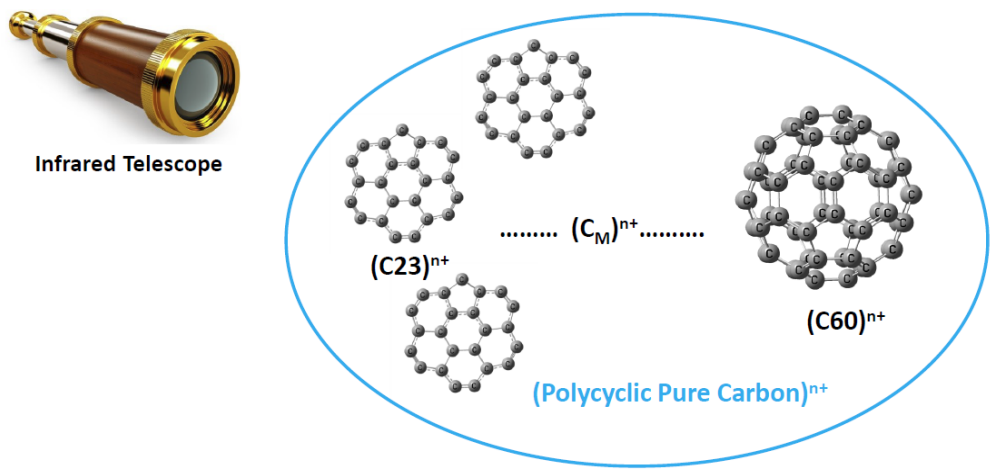

Figure 10, Infrared telescope may observe piled spectrum from neutral and ionized $(C_{23}),,,(C_M),,,(C_{60}), (C_{70})$ and so on.

7, CONCLUSION

Through the first principles quantum-chemistry calculation on deeply photo-ionized $(C_{23})^{n+}$, we could identify the astronomically observed infrared spectrum (IR) of carbon-rich hydrogen-poor planetary nebula Lin 49 and Tc 1.

(1) The first astronomical assumption is high speed particle attack, as like solar wind proton, on coronene-$(C_{24}H_{12})$ to create a carbon void and bring the quantum configuration change from $(C_{24}H_{12})$ to $(C_{23}H_{12})$ having two hydrocarbon pentagons combined with five hexagons.
(2) Another assumption is high energy photon attack on $(C_{23}H_{12})$ from the central star.
(3) There occurs deep photo-ionization on $(C_{23}H_{12})$. At a tenth step of $(C_{23}H_{12})^{10+}$, all hydrogen are decoupled from the molecule to be $(C_{23})$ having one carbon pentagon combined with six hexagons.
(4) There succeeded deeper photo-ionization on $(C_{23})^{n+}$. Infrared spectra of these molecules were calculated by molecular vibrational analysis.
(5) Calculated IR of $(C_{23})^{n+}$ (n=0, +1, and +2) could reproduce observed one of carbon-rich hydrogen-poor planetary nebula Lin 49 and Tc 1. We could identify 28 bands at wavelength from 6 to 28 micrometer, which include previously predicted bands of carbon fullerene (C60) at 7.0, 8.5, 17.4, 19.0, and 21.8 micrometer
(6) We found that a set of ionized fullerene $(C_{60})^{n+}$ (n=0, +1, and +2) can reproduce observed 10 bands.

It was concluded that calculated infrared spectrum of $(C_{23})^{n+}$ (n=0, +1, and +2) could reproduce observed bands very well.

Author profile
Norio Ota PhD.
Fellow and Honorable member, The Magnetics Society of Japan
Senior Professor, University of Tsukuba
Specialty: Material Science
Magnetic and Optical information devices

**Submit to arXiv.org    November    , 2018 by Norio Ota**

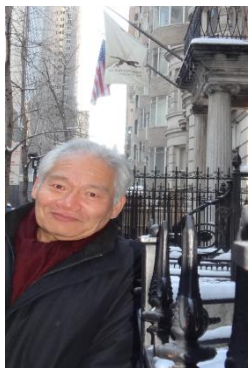

@New York